\documentclass{article}
\usepackage{graphicx}
\usepackage{xcolor}
\usepackage{listings}
\usepackage{epstopdf}
\usepackage{subfigure}
\usepackage{hyperref}
\hypersetup{colorlinks, citecolor=green, filecolor=black, linkcolor=blue, urlcolor=blue }
\usepackage{graphicx}
\usepackage{amsmath}
\usepackage[version=4]{mhchem}
\usepackage{siunitx}
\usepackage{longtable,tabularx}
\title{Accelerating High-Strain Continuum-Scale Brittle Fracture Simulations with Machine Learning}
\author{M. Giselle Fern\'andez-Godino, Nishant Panda, Daniel O'Malley, \\Kevin Larkin, Abigail Hunter, Raphael T. Haftka, and Gowri Srinivasan}

\begin{document}
    
    \maketitle
    
    \begin{abstract}
        
        Failure in brittle materials under dynamic loading conditions is a result of the propagation and coalescence of microcracks. Simulating this mechanism at the continuum level is computationally expensive or, in some cases, intractable. The computational cost is due to the need for highly resolved computational meshes required to capture complex crack growth behavior, such as branching, turning, etc. Typically, continuum-scale models that account for brittle damage evolution homogenize the crack network in some way, which reduces the overall computational cost, but can also neglect key physics of the subgrid crack growth behavior, sacrificing accuracy for efficiency. We have developed an approach using machine learning that overcomes the current inability to represent micro-scale physics at the macro-scale. Our approach leverages damage and stress data from a high-fidelity model that explicitly resolves microcrack behavior to build an inexpensive machine learning emulator, which runs in seconds as opposed to the high-fidelity model, which takes hours. Once trained, the machine learning emulator is used to predict the evolution of crack length statistics. A continuum-scale constitutive model is then informed with these crack statistics, speeding up the workflow by four orders of magnitude. Both the machine learning model and the continuum-scale model are validated against a high-fidelity model and experimental data, respectively, showing excellent agreement. There are two key findings. The first is that we can reduce the dimensionality of the problem, establishing that the machine learning emulator only needs the length of the longest crack and one of the maximum stress components to capture the necessary physics. Another compelling finding is that the emulator can be trained in one experimental setting and transferred successfully to predict behavior in a different setting. 
        
    \end{abstract}

    \section{Introduction}\label{sec:introduction}
    
    The existence, evolution, interaction, nucleation and coalescence of cracks is key to modeling strength and damage behavior in several brittle materials such as granite, concrete, metals, and ceramics~\cite{meyer2000crack, paliwal2008interacting, escobedo2014effect, huq2019micromechanics}. Ideally, predicting damage evolution in practical applications must account for the presence of cracks, but such simulations using finite and/or discrete element simulations are computationally very expensive. Including details about crack network evolution at the macro-scale can take thousands of processor hours. The high cost is due to the need for highly resolved meshes required to resolve the crack network discretely~\cite{rabczuk2013computational}. The computational expense is especially large when multiple simulations are needed, as in the case of optimization or uncertainty quantification. Current continuum-scale models use empirical models or statistical averaging techniques to inform damage~\cite{kachanov2013introduction}. However, detailed information about the material damage evolution is lost during this homogenization process~\cite{krajcinovic1987continuum,krajcinovic1995some}. 
    
    A possible solution to this problem is machine learning emulators, also known as surrogate emulators~\cite{queipo2005surrogate, fernandez2016review, fernandez2019issues}. These are built using data from expensive physical models achieving mainly two things: (i) they provide approximations that reduce computations by orders of magnitude and (ii) identify patterns in high dimensional data that are hidden to the human eye. Machine learning is usually most accurate when exposed to inputs that are similar to the training data. Therefore, machine learning emulators are a natural fit for problems that have a large and rich dataset for training and testing throughout the entire domain of interest. Another desirable ingredient is the availability of the necessary computational resources to obtain training and test data from simulations, which is often computationally intensive. 
    
    The combination of the recent growth in computation power along with the availability of a large amount of data laid the foundation for the resurgence of machine learning applications~\cite{marsland2015machine}. Machine learning emulators has been successfully used for handling classification~\cite{suthaharan2014big}, regression~\cite{huang2019application}, bridging scales~\cite{raissi2017physics,cheng2019bridging,wang2020deep}, and dimensionality reduction~\cite{cichocki2016tensor} problems. Machine learning has demonstrated to outperform humans in numerous tasks such as playing go~\cite{chen2016evolution}, driving cars~\cite{bojarski2016end}, and find hidden patterns~\cite{vasavi2018extracting}. Machine learning has made impressive advances in material science in recent years~\cite{schmidt2019recent}, in particular bridging scales in fracture mechanics~\cite{srinivasan2018quantifying, hunter2018reduced, moore2018predictive, schwarzer2019learning, mudunuru2019surrogate}.
    
    The physical problem of interest for this work is the study of the strength and damage behavior of beryllium under high strain dynamic loading conditions. We have developed a strategy to improve the strength model of a continuum-scale physical model using crack statistical information from the damage model of a high-fidelity model through machine learning in a way that can reduce simulation costs while maintaining accuracy. The current work takes into account the detailed information available in the high-fidelity model to account for damage evolution in the continuum-scale model through a machine learning algorithm. Furthermore, we reduced the dimensionality of the problem informed by the particular physics of the problem of interest and also investigating how much statistical information is required to achieve the desired match to experimental data. We use a long-short term memory (LSTM) recurrent neural network~\cite{gers1999learning,lipton2015learning} based machine learning emulator due to its outstanding prediction capability for time series data. 
    
    In addition to validating the machine learning emulator against high-fidelity simulations, we also validate the continuum-scale model against experimental data. Measuring crack lengths and stresses during experiments is not possible. Instead, experimentalists use VISAR (Velocity Interferometer System of Any Reflector) to measure the velocity at the middle rear of the target plate, which is presumed to be an indicator of the shock wave speed. Hence, the shock wave speed was the physical quantity used for validation. 
    We also investigate whether or not our machine learning model can be used outside of its training window. We show that for some conditions outside the training window, the machine learning model can perform well in comparison to an experiment. This work represents the first time that this workflow has succeeded in comparison to experiments. 
    
    In summary, the novelties and contributions of the present work are: 
    \begin{enumerate}
        \item The entire workflow of upscaling a continuum-scale model with a machine learning model (which was trained with a high-fidelity model) has been completed for the first time.
        \item A machine learning emulator is used to address brittle fracture in a high strain (instead of low strain) loading condition problem, including statistical information. 
        \item The relevant physics of the problem of interest was used to reduce its dimensionality which enabled the use of a simplified machine learning emulator.
        \item The machine learning emulator was validated outside of its training window for similar experimental conditions.
    \end{enumerate}
    
    This paper is organized as follows. Section~\ref{sec:problem} details the physical problem studied in this work. Section~\ref{sec:methods} includes a description of the high-fidelity model, HOSS, and of the continuum-scale model, FLAG. It also shows how damage is modeled in FLAG using crack statistics from HOSS. Section~\ref{sec:MLsec} describes how the physical variables to be trained using machine learning were chosen and shows detailed information on how the machine learning emulator was built. In Section~\ref{sec:Results}, the performance of the machine learning emulators is assessed, and plots are shown that validate the continuum-scale model upscaled using machine learning information against experiments. Finally, Section~\ref{sec:conclusion} summarizes this work, including a brief discussion of intended future work.
    
    \section{Problem of Interest} \label{sec:problem}
    
\begin{figure}[ht!] 
        \centering
        \includegraphics[width=0.9\columnwidth]{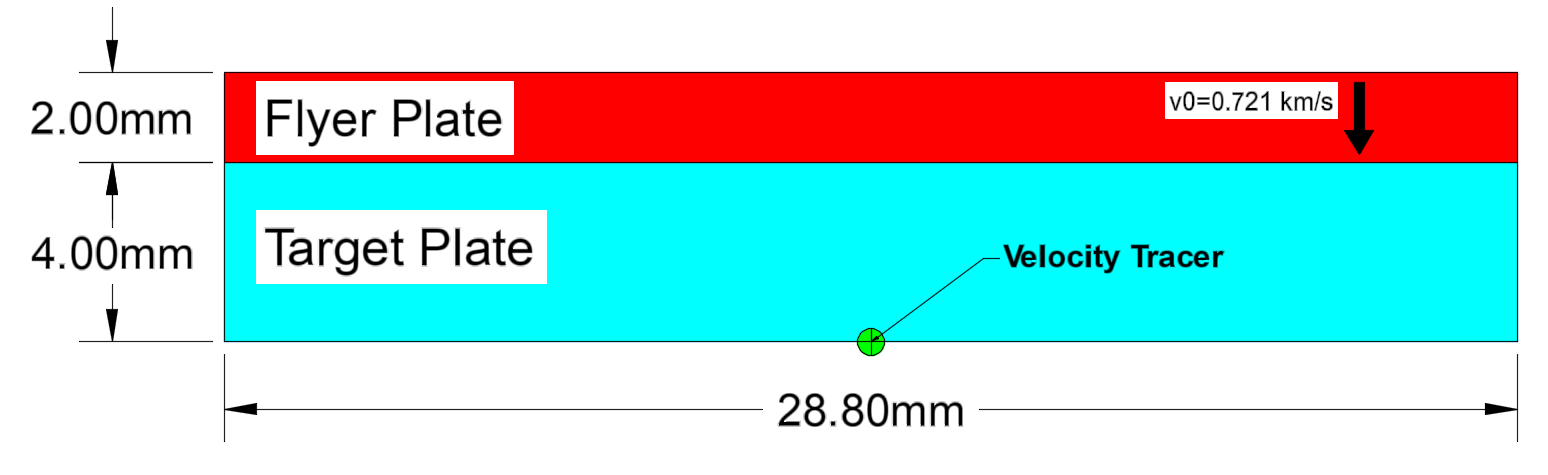} 
        \caption{Initial setup for the flyer plate test simulations. The flyer plate has an initial velocity of 0.721 $km/s$ and it is initially in contact with the target plate.}
        \label{FigSetup}
    \end{figure}

    In this paper, the physical problem of interest is the study of the strength and damage behavior of beryllium S200F under dynamic loading conditions. The simulated problem is a two-dimensional flyer plate impact against a target specimen. Both the flyer plate and the target are modeled as high-strength beryllium. Figure~\ref{FigSetup} is a schematic of the simulation setup. The sample has a width of $28.8mm$, the height of the target is $4 mm$, and the height of the flyer plate is $2 mm$. The flyer plate has an initial velocity of $0.721 km/s$, and the total simulation time is $1.2 \mu s$. The geometry and setup recreate the available flyer plate experiment 55-420 in Cady et al. .\cite{cady2011alamos} as closely as possible for validation purposes. A velocity tracer was placed at the middle rear of the target plate to measure the shock wave profiles for comparison with experiments that we use to validate our approach. 
    
    In fracture mechanics, we can define three main loading conditions: Mode I, Mode II, and Mode III. The difference between these loading conditions is the orientation of the force relative to the crack. Mode I, or opening, is when the loading is applied perpendicular to the crack. Mode II, or shear mode, refers to the loading applied in the in-plane direction. Finally, Mode III, or tearing mode, is when the shear stress is parallel to the leading edge of the crack. Other loading conditions can be represented as a combination of these. After the flyer plate impacts the target plate, the latter is subject to a strong compression that later becomes tension as the shock wave travels within the material and bounces against the borders of the plate. Due to this indirect uniaxial tensile load, we expect the problem of interest to be dominated by Mode I crack growth. The nature of fracture in flyer plate experiments leads to a non-homogeneous damage distribution where a concentrated region of damage forms across the midspan of the target plate. In contrast, the majority of the plate remains relatively undamaged. For more information on flyer plate experiments, the reader can refer to the references~\cite{cady2011alamos,cady2012characterization}.

    \section{Models}\label{sec:methods}
    
    \subsection{HOSS Model}\label{sec:HOSS}
    
    To generate the data needed to train our machine learning model, we choose the Hybrid Optimization Software Suite (HOSS) as our high-fidelity model~\cite{rougier2013lanl,knight2013lanl,knight2015hybrid}. HOSS is a finite-discrete-element method (FDEM) that can model the evolution of the microcrack network in high strain problems such as the ones we are interested in this work. HOSS simulates the evolution of discrete cracks within the crack network, which inform the machine learning emulator described later in Section~\ref{sec:ML}. HOSS is a hybrid multiphysics FDEM model that combines finite element techniques to describe the deformation of the material with discrete element-based transient dynamics, contact detection, and contact interaction solutions. It can account for both damage evolution, and catastrophic fracture or fragmentation. Detailed information about HOSS is out of the scope of this work, but a full description of HOSS FDEM is available in the literature \cite{munjiza1992discrete,munjiza1995combined,munjiza2004combined,munjiza2011computational,munjiza2015large} and its verification and validation~\cite{rougier2013benchmarking,rougier2014validation}.
    
    In HOSS FDEM, the discrete elements are divided into finite elements. The governing equations are conservation of mass, momentum, and energy along with Newton’s laws~\cite{munjiza1992discrete,munjiza1995combined,munjiza2004combined,munjiza2011computational,rougier2014validation,munjiza2015large}. HOSS uses an explicit central difference time integration scheme~\cite{rougier2004numerical}. In the HOSS FDEM, cracks can only form along the boundaries of the finite elements. To model mechanisms such as crack nucleation, coalescence, propagation, branching, reorientation, etc., tens to hundreds of finite elements along the length of each crack are necessary~\cite{munjiza2004combined}. Such a large number of finite elements results in gigabytes of data even for simulations involving laboratory sized samples with thousands of microcracks. The fine grids, combined with the explicit time integration scheme, can result in a need to utilize high-performance computing resources for an extended period to perform a HOSS simulation.
    
When HOSS models a pure tension problem, not every discrete element edge is oriented orthogonally to the applied load. For this reason, although globally opening failure dominates this problem, both shear and tearing modes occur at a local mesh element scale. Between the interface of any two finite elements, there are cohesive points modeled as springs. As the two elements undergo tensile load and are pulled apart, the springs within the interface are strained, and as a result, a small space between the elements develops. Similarly, for shear deformation, cohesive points can deform, allowing one element to slide relative to another. 
    
    For the flyer plate problem described in Section~\ref{sec:problem}, the total number of HOSS simulations available is 100, each having 480 time steps. The simulation time is $1.2 \mu s$, and the time step is $0.0025 \mu s$. To accurately simulate the crack dynamics, HOSS requires 31,000 elements, resulting in 23 GBs of data per simulation. The HOSS simulations used in this work required $160$ processor hours each ($2.5$ hours on $64$ processors). The HOSS computational burden is intractable for many real-world applications, such as optimization or uncertainty quantification, highlighting the need for machine learning emulators, such as surrogate or reduced-order models.
    
    HOSS is a deterministic model, which means that if HOSS is run using the same inputs, the output does not change if run multiple times. To obtain the statistical variability naturally existent in real-world materials, we impose a randomly generated distribution of initial cracks. For each simulation considered the target specimen initially has 200 cracks. The initial crack location $(x,y)$ within the target plate follows a uniform distribution in the horizontal coordinate ($x$), $x \sim U[0,4 mm]$, and in the vertical coordinate($y$), $y \sim U[0,28.8 mm]$, (see Figure~\ref{FigSetup}). The initial orientation of the cracks ($\theta$) also follows a uniform distribution $\theta \sim U[0^\circ,180^\circ)$. Note that an orientation of $0^\circ$ is equal to an orientation of $180^\circ$. That is why the latter is excluded from the interval. The initial crack lengths vary between $0.1mm$ and $0.3mm$ and are based on a power-law probability density function (PDF) that is inspired by the preexisting crack length distribution on materials~\cite{ignatovich2019power}. The power-law PDF limits ensure that at least two discrete elements exist along the crack. The initial position, orientation, and length of the preexisting cracks are changed randomly in every simulation. As stated before, to calculate the statistical crack evolution information, each HOSS simulation required a highly resolved mesh and took hours to complete on 64 processors. The PDF considered to generate the cracks length distribution on the target plate was the following power-law function:
    
    \begin{equation}\label{eq1}
    \textnormal{PDF }(x)=\frac{gx^{(g-1)}}{b^g-a^g},
    \end{equation}
    where $g=-3$, $a=0.1$, $b=0.3$, and $x$ is a vector in the range $[a,b]$. Note that $a$ and $b$ are the initial minimum and maximum crack length, respectively. Hence,
    
    \begin{equation}\label{eq2}
    \textnormal{PDF }(x)\approx \frac{x^{-4}}{321}.
    \end{equation}
    The cumulative density function (CDF) corresponding to this PDF is used to obtain samples. The CDF is given by
    
    \begin{equation}\label{eq3}
    \textnormal{CDF }(x)= \int_{a}^{x} \textnormal{PDF }(x) dx = \frac{x^{g}-a^g}{b^g-a^g} \approx \frac{1000-x^{g}}{963} .
    \end{equation}
    
    The CDF of Eq.~\eqref{eq3} is used to get a 200 sample crack length distribution for each simulation. Therefore given a random number $r$ in the interval $[0,1]$ each sample is obtained from
    
    \begin{equation}\label{eq4}
    x= [a^g + (b^g-a^g)r]^{1/g}\approx \frac{1}{\sqrt[3]{1000+963r}}. 
    \end{equation}
    
    For each of the 480 output files produced by a single HOSS simulation, we compute the lengths and orientations of every crack in the sample and compile the time-dependent crack PDF.
    
\begin{figure}[ht!] 
        \centering
        \includegraphics[width=0.9\columnwidth]{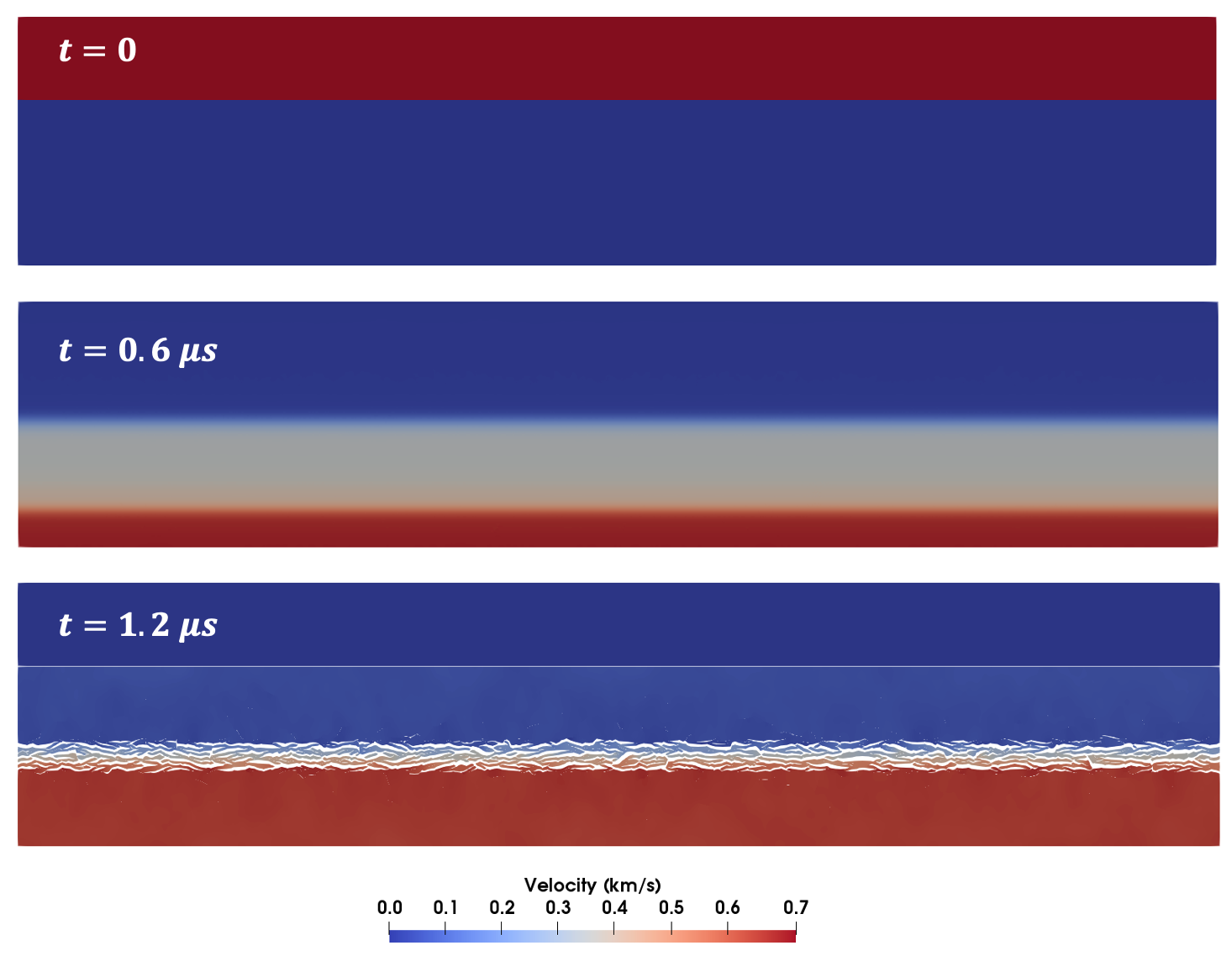} 
        \caption{Shock wave velocity at initial, intermediate and final times. The figures have been extracted from from HOSS simulations.}
        \label{fig:velocity}
    \end{figure}

\begin{figure}[ht!] 
        \centering
        \includegraphics[width=0.89\columnwidth]{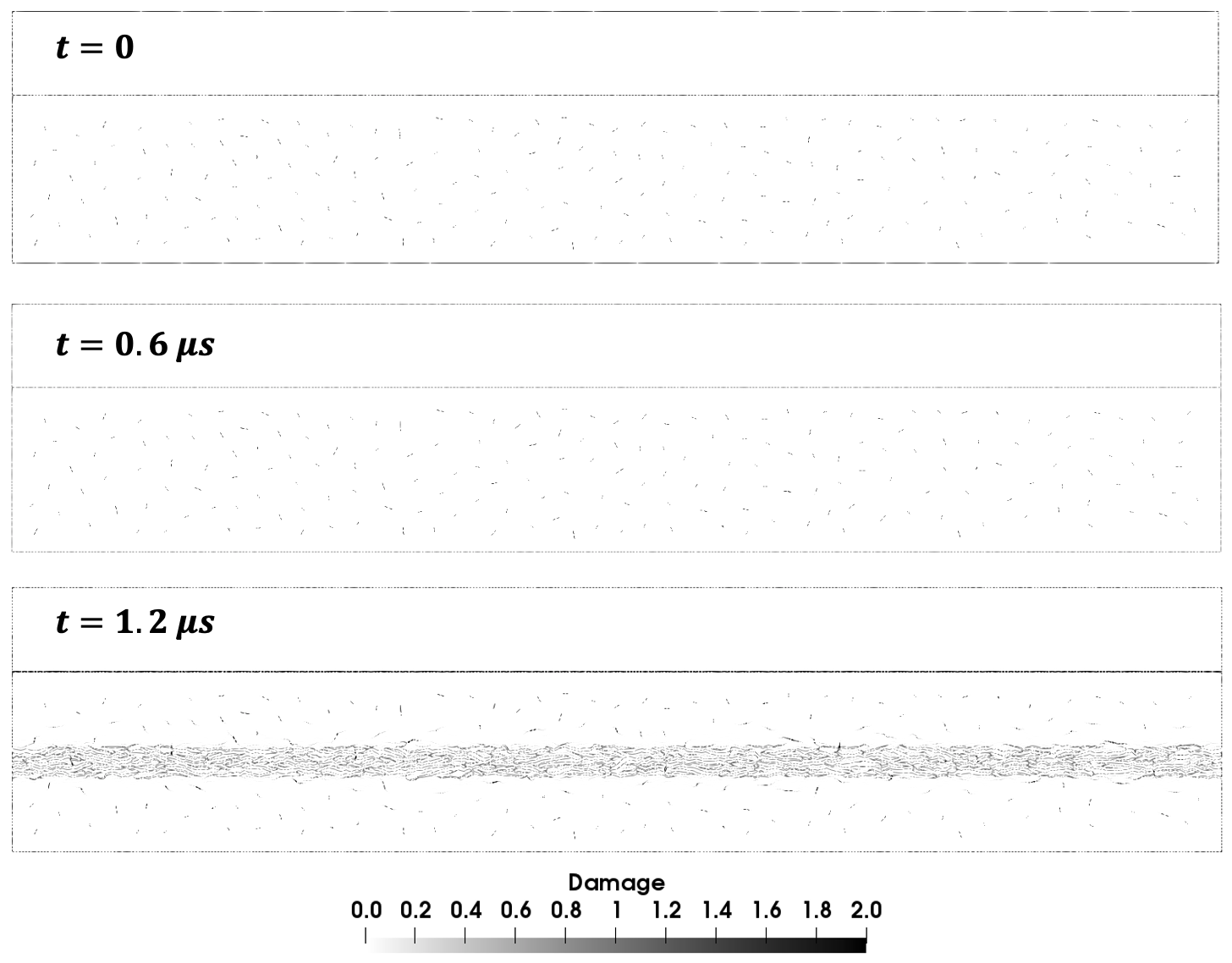} 
        \caption{Damage at initial, intermediate and final times. The figures have been extracted from HOSS simulations.}
        \label{fig:damage}
    \end{figure}

HOSS simulations of the flyer plate problem are visualized in Figure~\ref{fig:velocity} and Figure~\ref{fig:damage}. Figure~\ref{fig:velocity} shows the shock wave velocity at the initial time (t=0), at an intermediate time (0.6 $\mu s$) and at the final time (1.2 $\mu s$). Similarly, Figure~\ref{fig:damage} shows the damage at the initial time (t=0), at an intermediate time (0.6 $\mu s$ and at the final time (1.2 $\mu s$). Figure~\ref{fig:damage} shows the failure is produced in the middle section of the target plate, which is called a spall region. This is because the target plate is twice the width of the flyer plate. The target plate initially has 200 cracks, but additional cracks form in the spall region as the simulation proceeds. At the final time, the number of cracks is lower due to crack coalescence.

    \subsection{FLAG Model}\label{sec:FLAG}
    
    \subsubsection{Description}
    
    For this work, we choose to utilize the ML-informed constitutive model into a hydrodynamics code called FLAG ~\cite{tonks2010mesoscale,tonks2012comparison,fung2013ejecta}. FLAG is a fully parallel multiphysics model developed and maintained by Los Alamos National Laboratory~\cite{Burton1992FLAG, Burton1994aFLAG, Burton1994bFLAG}. FLAG is an unstructured polyhedral cell-centered or staggered-grid finite volume element model, coupled with arbitrary Lagrangian-Eulerian (ALE) remapping~\cite{Burton2018ALE} or adaptive mesh refinement (AMR) mesh optimization techniques, which allow for the simulation of complex multidimensional dynamics problems. FLAG has a large array of physics packages such as the various equation of state models as well as material strength, plasticity, and damage models that make FLAG especially useful for simulating shock physics problems. Previous researchers have used FLAG to study impact cratering~\cite{Caldwell2018Crater}, ejecta formation and transport~\cite{fung2013ejecta}, flyer plate impact experiments~\cite{Cooley2014fyler}, shock-driven multiphase instabilities~\cite{Black2017shock}, and detonation shock dynamics~\cite{Aida2013shock}. We aim to use FLAG to simulate damage in flyer plate impact experiments using a statistical crack growth model, informed by crack data generated using machine learning techniques.
    
    The beryllium flyer plate experiment depicted in Figure \ref{FigSetup} is simulated in FLAG using a 2D Lagrangian framework. The flyer plate and target plate were meshed using 10368 and 20352 quad elements (zones), respectively. Running a single flyer plate simulation with FLAG requires $0.1$ processor hours. Hence the computational cost of the high-fidelity model, HOSS, is 1600 times higher than the computational cost of FLAG. The left and right edges of the flyer and target plates are constrained to move only in the vertical direction, and a sideline boundary condition is placed between the plates to avoid the interpenetration of the two meshed objects. The Gruneisen analytical equation of state~\cite{Fredenburg2015EOS} and the Steinburg-Guinan plasticity model~\cite{Steinberg1980plas} are used to simulate the material properties of beryllium. The parameters for these models can be found in Tables \ref{table:EOSpara} and \ref{table:SGpara}, respectively. The damage model used to simulate crack propagation in the target uses a modified version of the statistical effective moduli model of Ju and Chen~\cite{ju1994effective}. This model relies on crack length and orientation data as well as the maximum tensile stress at each time step to approximate the effective compliance of each zone within the cracked material. In this work, this data is extracted from the HOSS model. The damage model and its connection to machine learning are discussed in more detail in Section \ref{sec:Coupling}. The velocity at the rear center of the target plate was recorded for comparison with the high-fidelity model and experimental VISAR data.
    
    \begin{table}[h]
        \caption{\label{table:EOSpara}Gruneisen EOS model parameters for beryllium}
        
        \centering
        \begin{tabular}{ccc}
            \hline
            Parameter&Description&Value\\
            \hline
            $\rho_0$ & reference density & 1.845$g/cm^3$\\
            $c$ & bulk sound speed & 0.799$cm/\mu s$\\
            $s_1$ & linear Hugoniot coefficient & 1.13\\
            $\gamma_0$ & Gruneisen $\gamma$ at initial density & 1.11\\
            $a$ & ramp parameter & 0.16\\
            $t_0$ & reference temperature & 293$K$\\
            $c_v$ & specific heat & 1.82$J/gK$\\ 
            \hline
        \end{tabular}
    \end{table} 
    
    \begin{table}[h]
        \caption{\label{table:SGpara}Steinberg-Guinan model parameters for beryllium~\cite{Steinberg1996plas}}
        
        \centering
        \begin{tabular}{ccc}
            \hline
            Parameter&Description&Value\\
            \hline
            $\rho_0$ & reference density & 1.845$g/cm^3$\\
            $G_0$ & initial shear modulus & 1.51$Mbar$\\
            $Y_0$ & initial flow stress & 0.0033$Mbar$\\
            $Y_{max}$ & max work hardening & 0.0131$Mbar$\\
            $\beta$ & work hardening parameter & 26\\
            $n$ & work hardening exponent & 0.78\\
            $A$ & pressure dependence multiplier & 0\\
            $B$ & temperature dependence multiplier & 0\\
            $q_y$ & flow stress pressure dependence factor & 1.0\\
            $f_g$ & melt shaping for shear modulus & 0\\
            $f_y$ & melt shaping for flow stress & 0\\
            $\rho_{0s}$ & crushed-up density & 1.845$g/cm^3$\\
            \hline
        \end{tabular}
    \end{table}
    
    \subsubsection{Validation}
    
\begin{figure}[ht!] 
        \centering
        \includegraphics[width=0.8\columnwidth]{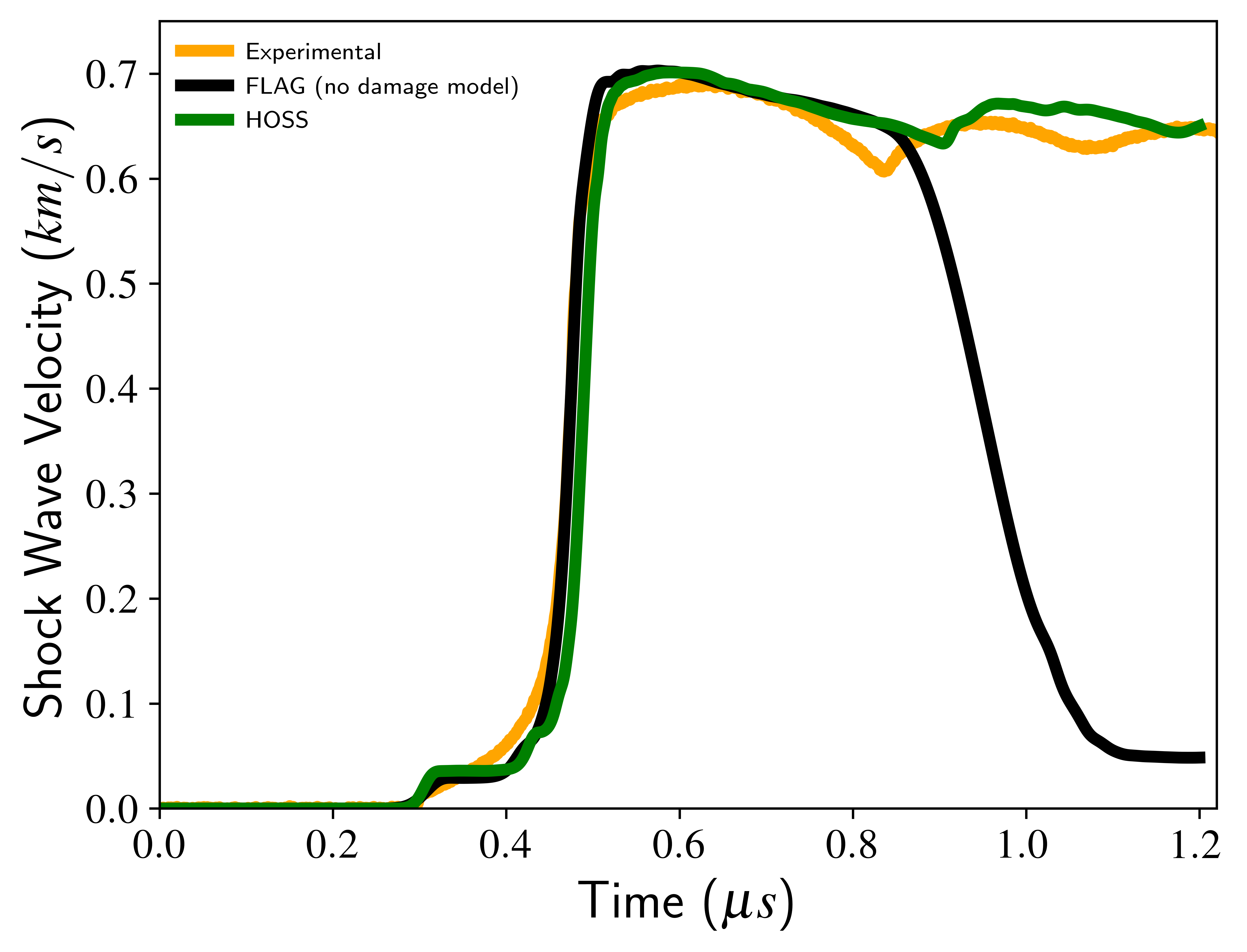} 
        \caption{Evolution of the shock wave velocity at the middle rear of the target plate.}
        \label{fig:visar1}
    \end{figure}

    Figure~\ref{fig:visar1} shows the shock wave velocity profiles at the middle rear of the target plate as a function of time. As the Figure shows, the experiment (orange curve) captures a shock wave velocity oscillation due to the shock reflection between the rear of the target plate and the spall region. The spall region is shown clearly in Figure~\ref{fig:damage} at $1.2 \mu s$. This oscillation is not well captured in FLAG simulations if damage models are turned off (black curve) since cracks, and hence the development of a spall region, are not taken into account. By contrast, HOSS simulations (green curve) simulate each crack individually so that they can reproduce the experimental behavior.
    
    \subsubsection{Upscaling}\label{sec:Coupling}
    
    As shown in Figure~\ref{fig:visar1}, without a damage model, FLAG can accurately simulate the initial shock wave that propagates through the target plate (until approximately 0.8$\mu s$). However, the subsequent pull-back signals, from the rebounding shock wave, are not captured because the model does not account for the damage region across the midspan of the target plate. Hence various upscaling procedures must be developed for approximating damage caused by microcracks and correcting the stresses predicted by FLAG.
    
    We take a statistical approach to approximating the damage caused by crack evolution in the target plate utilized in FLAG. In such a framework, Vaughn et al.~\cite{vaughn2019statistically} developed a modified version of the effective moduli model of Ju and Chen~\cite{ju1994effective} that approximates the compliance of the damaged material as follows
    
    \begin{equation}
    \boldsymbol{S}(t)=\boldsymbol{S^0}+\boldsymbol{S^*}(t)
    \label{eqSeff}
    \end{equation}
    
    \noindent where $\boldsymbol{S^0}$ is the pristine elastic compliance tensor, which in two dimensions and for isotropic materials is defined as 
    
    \begin{gather} \label{elastic}
    \boldsymbol{S^0} = \frac{(1+ \nu)}{E} 
    \begin{bmatrix}
    1-\nu & -\nu & 0 \\
    -\nu & 1-\nu & 0 \\
    0 & 0 & 2
    \end{bmatrix},
    \end{gather}
    
    \noindent where $\nu$ is Poisson's ratio, and $E$ is Young's modulus of the material. $\boldsymbol{S^*} (t) $ is a compliance tensor correction that is added to the elastic compliance tensor to account for microcrack effects at time $t$ which has the following expression
    
    \begin{equation}
    \boldsymbol{S^*} (t)=\bar{L}\frac{\pi(1-\nu^2)n(t)}{E}\int_{\alpha}\int_{\theta}{a^2}{\boldsymbol{M_0}f(a,t)f(\theta,t)}d\theta da
    \label{eqSstar}
    \end{equation}
    
    \noindent where $f(a,t)$ and $f(\theta,t)$ are the PDFs of crack length $a$, and orientation $\theta$ at time $t$, and $n(t)$ is the number of cracks at time $t$. The coordinate transformation tensor $\boldsymbol{M_0}$ relates the local coordinates of the cracks to the global coordinate system. The interested reader can refer to~\cite{ju1994effective} for a detailed derivation of $\boldsymbol{M_0}$. The dimensionless parameter $\bar{L}$ added by Vaughn et al.~\cite{vaughn2019statistically} accounts for the finite length of the material domain and is determined using the following relation
    
    \begin{equation}
    \bar{L}=\left(\frac{L}{L-\bar{a}}\right)^p
    \label{eqLbar}
    \end{equation}
    
    \noindent where $L$ is the width of the target plate, $\bar{a}$ is the orthogonal projection to the loading direction of the longest crack in the domain, and $p$ is the domain rescaling factor exponent~\cite{vaughn2019statistically}. In this work, we use $p=1$, which has been calibrated with high rate experiments on Be~\cite{larkin2020scale}.
    
    This stress-based damage evolution scheme developed by Larkin et al. 2020~\cite{larkin2020scale} is employed to evaluate the damage level of each material zone. The stress-based scheme was developed to initiate damage when tension occurs (i.e., Mode I failure). Therefore, the maximum tensile stress from the statistical data set is correlated with a compliance correction tensor generated at the same time step. The appropriate compliance correction is identified for each zone by comparing a stress estimate to the maximum stress statistics. The maximum tensile stress is extracted from HOSS simulations, and it is the maximum tensile stress in any zone of the target plate. Then the stress value is updated using the identified effective moduli using the following relations.
    
    \begin{equation}
    \boldsymbol{\sigma}=\boldsymbol{C}:\boldsymbol{\gamma} 
    \label{eqStress}
    \end{equation}
    
    \begin{equation}
    \boldsymbol{C}=\boldsymbol{(S)^{-1}}
    \label{eqC}
    \end{equation}
    
    \noindent where $\boldsymbol{\sigma}$ is the corrected stress, $\boldsymbol{\gamma}$ is the elastic stain, and $\boldsymbol{C}$ is the effective stiffness tensor. 
    
    In our past work, crack length and orientation data, as well as maximum tensile stress data, were extracted directly from HOSS simulations~\cite{vaughn2019statistically,larkin2020scale}. The crack data was used to generate the distributions $f(a,t)$ and $f(\theta,t)$ in Eq.~\eqref{eqSstar}, which was then used to the effective moduli model in FLAG at each time step. However, in this work, we use machine learning to emulate the tensile stresses and the crack length statistics for the distributions $f(a,t)$. 
    
    \section{Machine Learning} \label{sec:MLsec}
    
    \subsection{Dimensionality Reduction and Training Data} \label{sec:DR}
    
\begin{figure}[!htbp]
        \centering 
        \subfigure[$t=0$\label{fig:Or_PP1}]
        {\includegraphics[width=2.3in,height=1.8in]{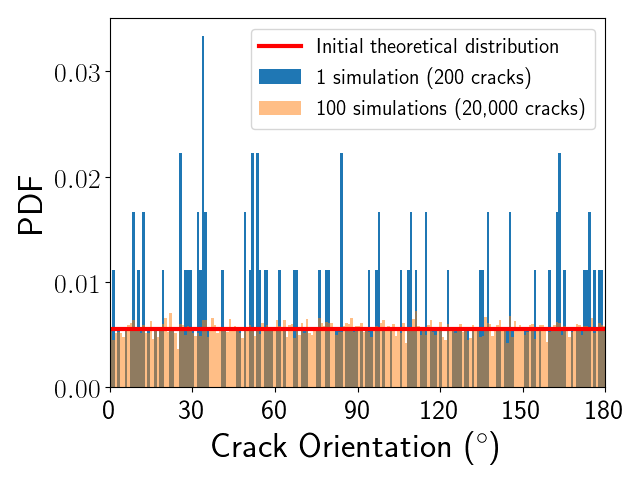}}\hspace{0.8em}
        \subfigure[$t=0.4 \mu s$\label{fig:Or_PP2}]
        {\includegraphics[width=2.3in,height=1.8in]{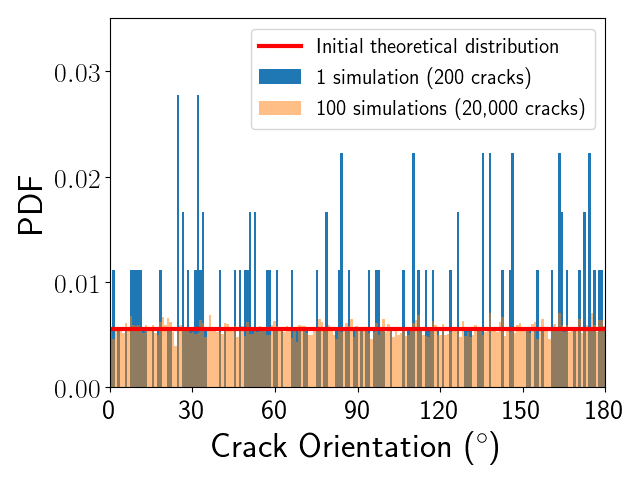}}\\ 
        \subfigure[$t=0.8 \mu s$\label{fig:Or_PP3}]
        {\includegraphics[width=2.3in,height=1.8in]{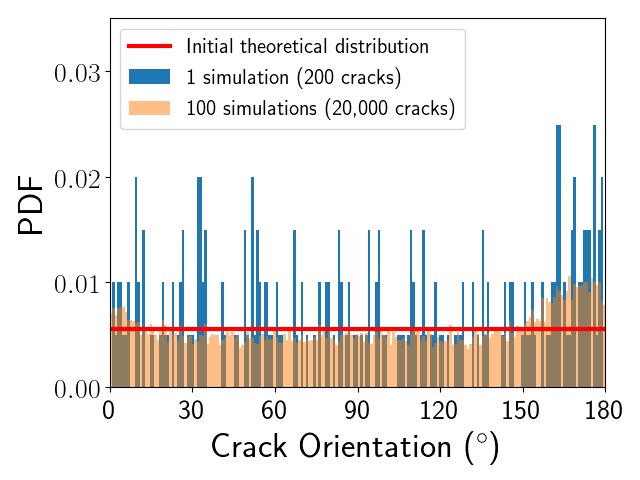}}\hspace{0.8em}
        \subfigure[$t=1.2 \mu s$\label{fig:Or_PP4}]
        {\includegraphics[width=2.3in,height=1.8in]{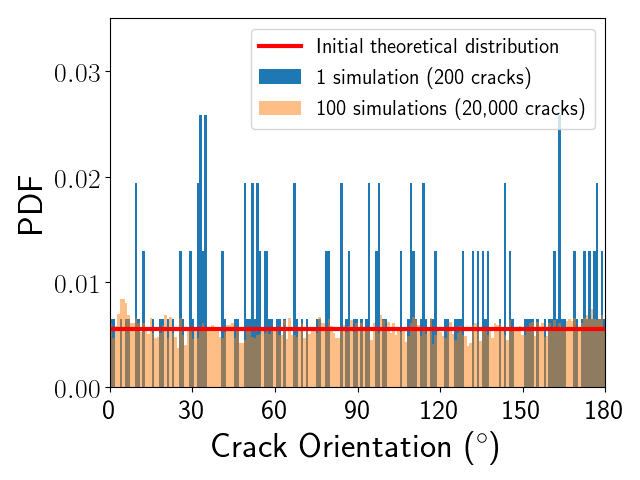}}
        \caption{Evolution of the crack orientation distribution in the interval $[0,180 ^{\circ})$ for one HOSS simulation (blue histogram) and for 100 HOSS simulations (orange histogram). The initial distribution (continuous red line) remains dominant throughout the simulation. \label{FigOrSimDist}}
    \end{figure}

    The first thing investigated was the effect of the evolution of the crack orientation on the overall macro-scale response. The HOSS statistics show that overall the crack orientations trend toward an orientation that is orthogonal to the loading conditions, however the change from the initial distribution is not dramatic. Figure~\ref{FigOrSimDist} shows the crack orientation PDF for one HOSS simulation (blue histogram), the crack orientation PDF for 100 HOSS simulations (orange histogram), and the initial crack orientation distribution (continuous red line). Figure~\ref{fig:Or_PP1} corresponds to the initial time ($t=0$), Figure~\ref{fig:Or_PP2} and Figure~\ref{fig:Or_PP3} correspond to the intermediate times $0.4 \mu s$ and $0.8 \mu s$, respectively, and Figure~\ref{fig:Or_PP4} corresponds to the final time ($1.2 \mu s$). At the initial time, the 200 crack orientations vary between $0$ and $180^{\circ}$. When the material failure occurs ($\approx 0.8 \mu s$), we observe crack nucleation (generation of approximately 150 new cracks per simulation), and the orientation of the nucleated cracks is mostly horizontal as Figure~\ref{fig:Or_PP3} shows. Because cracks coalesce at later times, at the final time, we observe a crack orientation distribution similar to the initial distribution (Figure~\ref{fig:Or_PP4}). As can be seen in Figure~\ref{FigOrSimDist}, regardless of the time, the initial distribution of crack orientation maintains throughout the entire time.

\begin{figure}[ht!] 
        \centering
        \includegraphics[width=0.8\columnwidth]{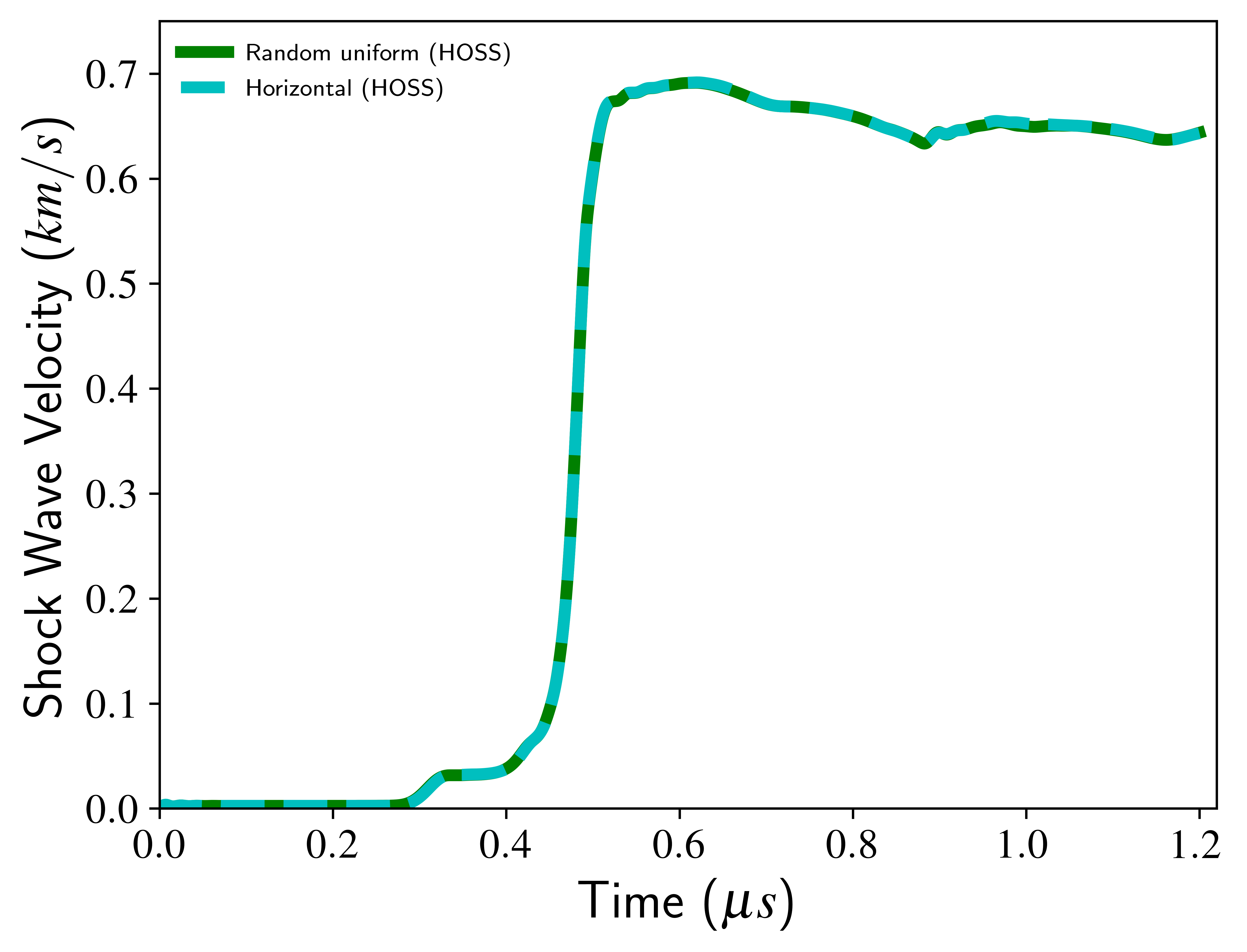} 
        \caption{Evaluation of the shock wave velocity at the middle of the rear of the target plate for two cases: i) random uniform orientation distribution and ii) horizontal orientation. As the figure shows the difference between the two cases is negligible. }
        \label{fig:horizontal}
    \end{figure}

    We run HOSS utilizing two initial orientation distributions: i) the randomized uniform initial distribution ($\theta \sim U[0^\circ,180^\circ)$), and ii) a distribution in which all the cracks are oriented orthogonal to the loading direction (horizontal in this setup corresponds to $\theta=0$ or $\theta=180^{\circ}$). The latter case is representative of a worst-case scenario, in which all the cracks are best oriented for Mode I opening. The resulting velocity profiles calculated with HOSS are shown in Figure~\ref{fig:horizontal}, which shows that there is no discernible difference in the macro-scale response.
    
    As Figure~\ref{fig:horizontal} shows, we determined that the influence of $f(\theta)$ was negligible compared with $f(a)$ and the tensile stresses (Eq.~\ref{eqSstar}). Then, as a result of this study, $f(\theta)$ was not trained using machine learning and only the initial $f(\theta)$ distribution was used to calculate Eq.~\eqref{eqSstar} at each time step, i.e. the distribution was not evolved.
    
\begin{figure}[!htbp]
        \centering 
        \subfigure[$t=0$\label{fig:PP1}]
        {\includegraphics[width=2.3in,height=1.8in]{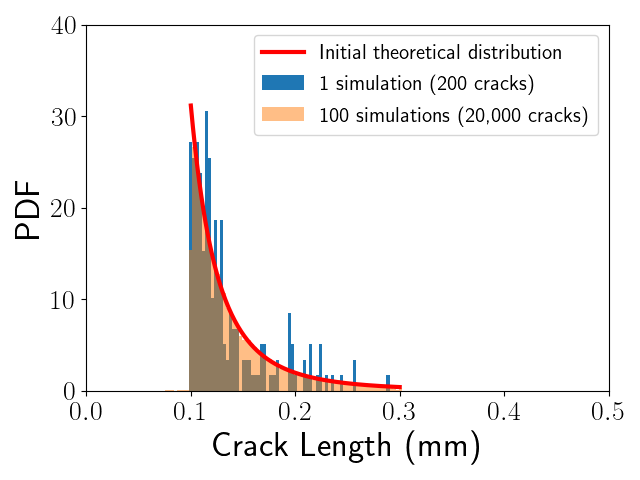}}\hspace{0.8em}
        \subfigure[$t=0.4 \mu s$\label{fig:PP2}]
        {\includegraphics[width=2.3in,height=1.8in]{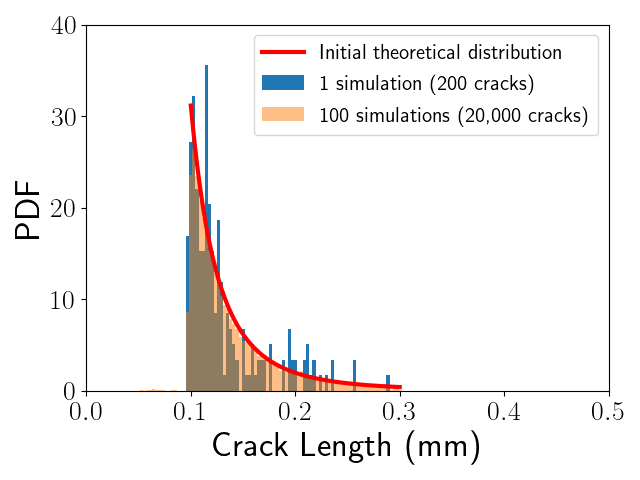}}\\ 
        \subfigure[$t=0.8 \mu s$\label{fig:PP3}]
        {\includegraphics[width=2.3in,height=1.8in]{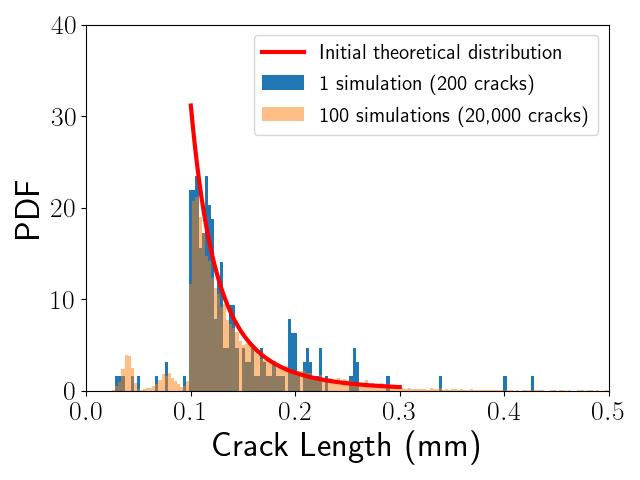}}\hspace{0.8em}
        \subfigure[$t=1.2 \mu s$\label{fig:PP4}]
        {\includegraphics[width=2.3in,height=1.8in]{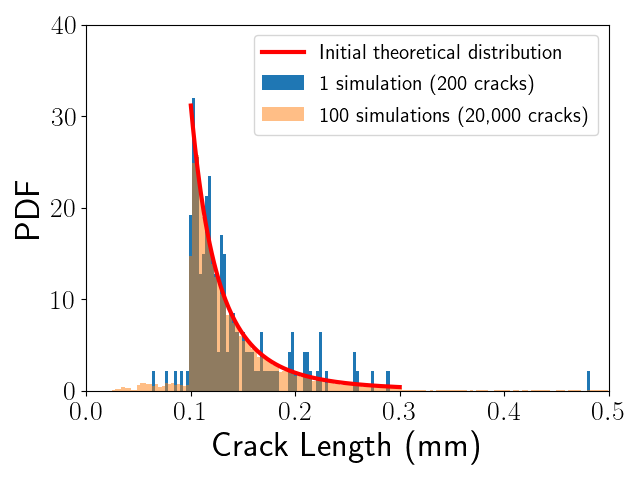}}
        \caption{Evolution of the crack length distribution in the interval $[0,0.5] mm$ for one HOSS simulation (blue histogram) and for 100 HOSS simulations (orange histogram). Near the time when the material fails ($ \approx 0.8 \mu s$) we can see substantial crack nucleation and crack length growth. The initial distribution (continuous red line) remains dominant throughout the simulation. \label{FigSimDist}}
    \end{figure}

    The second thing investigated was the effect of the evolution of the crack length on the overall macro-scale response. Figure~\ref{FigSimDist} shows the crack length PDF for one HOSS simulation (blue histogram), the crack length PDF for 100 HOSS simulations (orange histogram), and the initial crack length distribution (continuous red line). Figure~\ref{fig:PP1} corresponds to the initial time ($t=0$), Figure~\ref{fig:PP2} and Figure~\ref{fig:PP3} correspond to the intermediate times $0.4 \mu s$ and $0.8 \mu s$, respectively, and Figure~\ref{fig:PP4} corresponds to the final time ($1.2 \mu s$). At the initial time, the 200 crack lengths vary between $0.1mm$ and $0.3mm$. Until $0.8 \mu s$, the maximum crack length remains constant. When the material failure occurs ($\approx 0.8 \mu s$), we observe crack nucleation (generation of approximately 150 new cracks per simulation) and also crack growth beyond the initial maximum crack length. At the final time, we observe only around 160 cracks, because cracks coalesce at later times. As can be seen in Figure~\ref{FigSimDist}, regardless of the time, the initial distribution of cracks maintains and only a few cracks per simulation grow (less than five cracks) larger than 0.3 mm. For clarity, the plots are shown on the same scale. However, after $\approx 0.8 \mu s$, the longest crack has a length of $\approx 28.8 mm$, which is the width of the target plate. Nevertheless, Figure~\ref{fig:PP3} and Figure~\ref{fig:PP4} show some of the cracks with a length that exceeds $0.3 mm$. 
    
    While emulating the evolution of maximum tensile stresses can be challenging, emulating the evolution of each crack is a particularly difficult task for the following reasons:
    \begin{enumerate}
        
        \item The number of cracks changes as a function of time due to nucleation and coalescence of cracks. The major challenge is that the input/output of the machine learning emulator should allow variable size at each time step.
        
        \item While the initial distribution of crack lengths seems to be dominant throughout the entire simulation, the distribution changes significantly in the tails. Small new cracks nucleate, and a few cracks grow substantially due to coalescence. This behavior presents a challenge for the machine learning emulator because it is easier to learn typical behavior rather than outlier behavior; hence it may not learn the behavior of the rare, long cracks. In the physical problem studied in this work, these longer cracks determine the failure of the material, hence are of high importance. Machine learning needs to be informed about this particular aspect of the physics of the problem.
        
        \item The number of training examples to simulate a full-time series starting from the initial conditions need to be sufficiently rich. In this case, we have only 100 simulations, which, even if the initial conditions do not dramatically change the output, are not enough to obtain reasonable predictions for the full distribution.
        
    \end{enumerate}

    Taking into account the reasons enumerated above, an approach that allows the machine learning emulator to keep constant the input/output size, that prioritizes the information associated with the longest cracks at a given time and that reduces the dimensions needed to train was necessary. Besides not evolving the crack orientations, to get a further dimensionality reduction, we decided to focus only on the longest crack evolution.

    This led to the following update of Eq.~\eqref{eqSstar}. The crack orientations PDF $f(\theta,t)$ was replaced by the initial crack orientations PDF $f(\theta,t=0)$. The PDF $f(a,t)$ in Eq.~\ref{eqSstar} was replaced by $f_{\text{LC}}(a,t)$ which is the distribution of crack lengths at the initial time where only the evolution of the length of the longest crack (LC) is updated. Hence, Eq.~\eqref{eqSstar} is modified as follows
    
    \begin{equation}
    \boldsymbol{\widehat{S}^*} (t)=\bar{L}\frac{\pi(1-\nu^2)n}{E}\int_{a}\int_{\theta}{a^2}{\boldsymbol{M_0}f_{\text{LC}}(a,t)f(\theta,t=0)}d\theta da.
    \label{eqSstar_mod}
    \end{equation}
    
\begin{figure}[!htbp] 
        \centering
        \subfigure[Length of the longest crack as a function of time extracted from 100 HOSS simulations.\label{fig:100Sim1}]
        {\includegraphics[width=0.45\columnwidth]{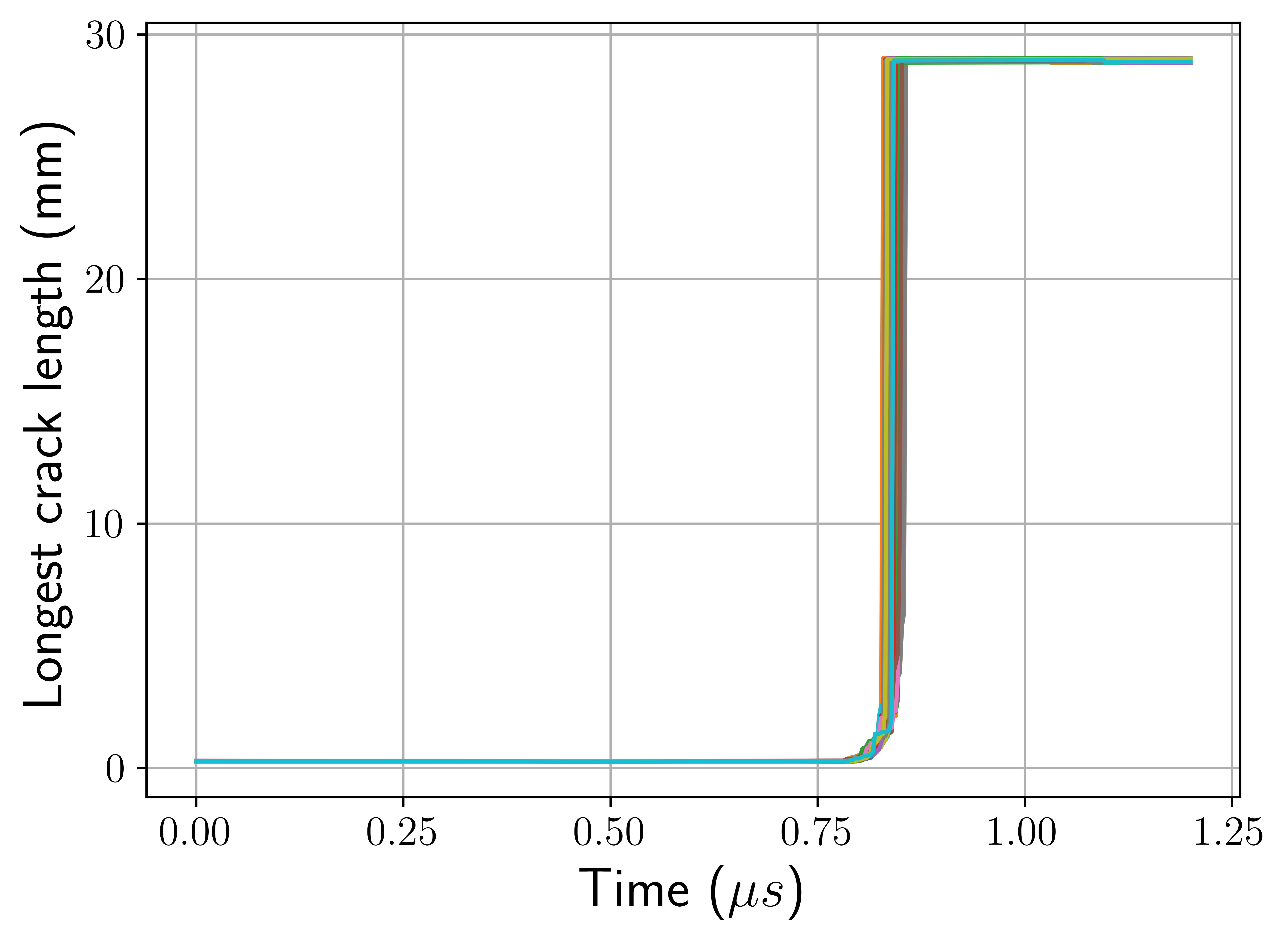} } 
        \hspace{2em}%
        \subfigure[Maximum tensile stress as a function of time extracted from 100 HOSS simulations.\label{fig:100Sim2}]
        {\includegraphics[width=0.45\columnwidth]{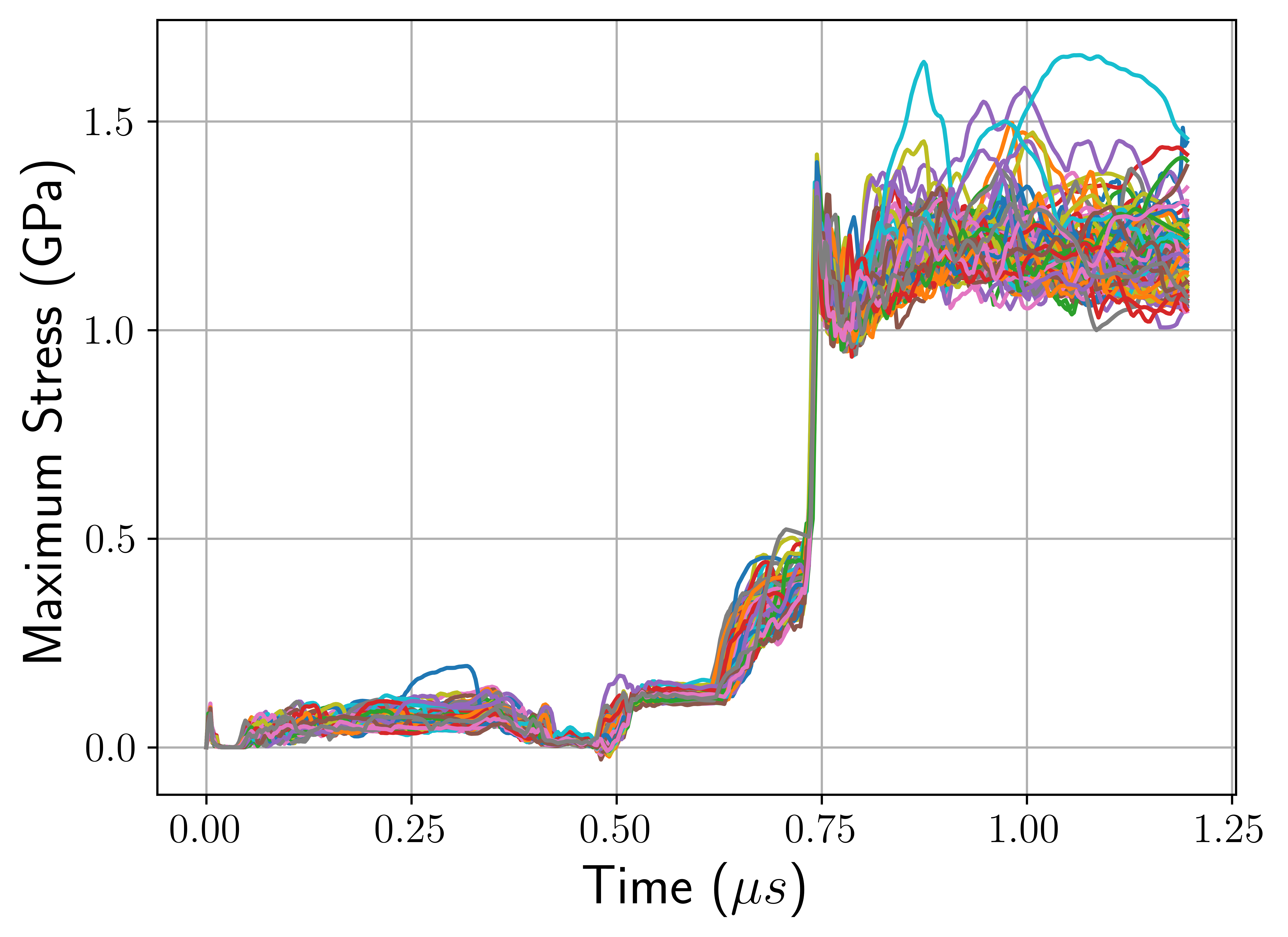} } 
        \caption{Physical variables to be used to train and test the machine learning emulator. \label{fig:100Sim}}
    \end{figure}

    Figure~\ref{fig:100Sim} shows the physical quantities chosen to train the machine learning emulator as a function of time. Figure~\ref{fig:100Sim1} shows the evolution of the length of the longest crack. As can be seen in the Figure, the higher variability from simulation to simulation occurs close to 0.8 $\mu s$. Figure~\ref{fig:100Sim2} shows the maximum tensile stress as a function of time. This quantity shows more variability than the length of the longest crack, especially after 0.8 $\mu s$. Having more variability generally allows the machine learning emulator to be more predictive. 
    
    \subsection{Baseline Emulator}
    
    It is a common practice when using machine learning emulators to compare their performance to a simple or baseline emulator. This sanity check assesses if the machine learning emulator is learning enough from the data and if it can perform better than a simple emulator. In this work, the mean value of the physical variables (i.e., the length of the longest crack and the maximum tensile stress as a function of time) of all the available training simulations at each time step was chosen as baseline emulator. Let us assume that we are interested in predicting the $j$th physical variable, $v_j$, then the baseline emulator for $v_j$ is computed as

    \begin{equation}
    \bar{v}_j (t)= \frac{1}{N}\sum_{i=1}^N{v_{ij} (t)}
    \label{eq:mean}
    \end{equation}
    where $v_{ij}$ is the $i$th simulation data of the variable $v_j$, $N$ is the total of simulations available and $t$ is time. From now, on we will refer to this simple emulator as baseline emulator. 
    
    \subsection{Machine Learning Emulator}\label{sec:ML}
    
    We seek to build a machine learning emulator that can predict two quantities of interest, namely the length of the longest crack and the maximum tensile stress as a function of time. We utilized a total of 100 simulations. Each simulation gives us the length of the longest crack and the maximum tensile stress for 480 time steps (see Figure~\ref{fig:100Sim}). We use 70 simulations for training the model and 30 for testing the model. These simulations were run varying the initial crack location, orientation, and length to obtain statistical variation. In the length of the longest crack and the maximum tensile stress obtained from the simulations, the variability from one simulation to the next is relatively small. This small variation is the result of modeling the same experiment while changing only the initial crack, location, orientations, and lengths from simulation to simulation to obtain statistical variation. In other words, the initial defect density does not have significant changes in the overall material response. This behavior supports the statement that the preexisting crack network used is realistic. The drawback of this small variability is that because we are predicting from the initial time step (initial condition), the model must be sensitive to small variations.

    \subsubsection{Training}
    
\begin{figure}[ht!] 
        \centering
        \includegraphics[width=0.9\columnwidth]{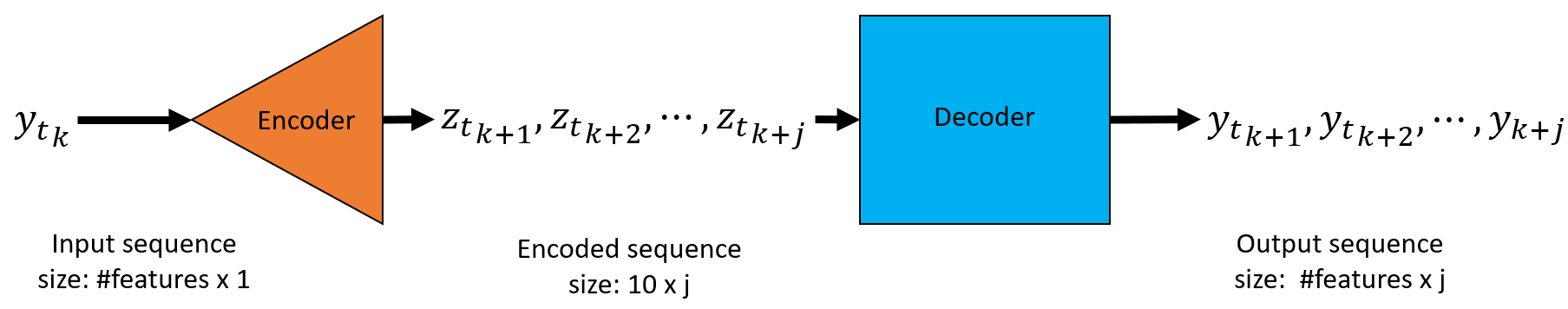} 
        \caption{Encoder-decoder model to sequence output. The subscript $t$ refer to the timestep considered. The network takes the input ${y_t}_k$ of size $2 \times 1$ in our case (two features) corresponding to the timestep $k$ and gives predictions in the timestep interval [$k+1,k+2,\dots,k+j$]. The encoder gives a sequence $z_t$ of size $j=100$, where each element has size $10 \times 1$ which is decoded to the output sequence of size $2 \times 100$.}
        \label{fig:network}
    \end{figure}

    The emulator was trained using Keras Python~\cite{chollet2015keras} Long Short Term Memory (LSTM) neural network. LSTMs often perform well at predicting sequences and time series~\cite{brownlee2018deep}. Our emulator receives two inputs: the initial length of the longest crack and the initial maximum tensile stress. The output consists of predictions of the length of the longest crack and maximum stress at later times. During training, the emulator predicts 100 future values of each of these quantities, and the loss function compares these predictions against the training data. The structure of the machine learning model as an encoder-decoder model is shown schematically in Figure~\ref{fig:network}.
    
    The network was trained for 400 epochs (i.e., the training process looked at each input/output pair in the training set 400 times). The loss function was the mean squared error (MSE) between the emulator predictions and training data. The Adam~\cite{kingma2014adam} variation of stochastic gradient descent was used to perform the optimization with a batch size of 14 (i.e., each stochastic estimate of the gradient used 14 input/output pairs).
    
    \subsubsection{Prediction}

\begin{figure}[ht!] 
        \centering
        \includegraphics[width=0.9\columnwidth]{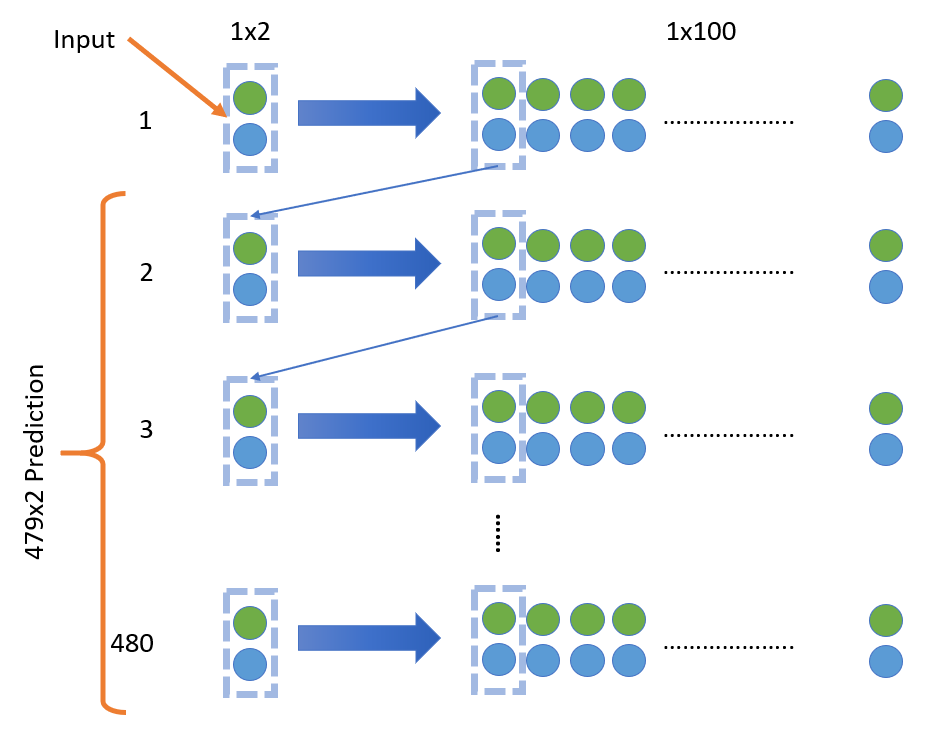} 
        \caption{Prediction process for a single input. The neural network built is used 480 times to obtain predictions for the 480 time-step interval of interest for each variable.}
        \label{fig:prediction}
    \end{figure}

    After successfully training the model, we use it in a predictive mode. While the model predicts 100 time steps at a time, we found better performance using only the values of the two physical variables (i.e., the length of the longest crack and the maximum tensile stress) at the next time step. This prediction then feeds into the network to calculate the next time step. This process repeats until we predict the entire sequence, i.e., 480 time steps. The final prediction includes the initial time step along with the 479 predictions (total size of $ 480 \times 2$ -- see Figure~\ref{fig:prediction}).

    \section{Results} \label{sec:Results}
    
    \subsection{Performance}
    
    To assess the performance of our models, we use the range averaged mean RMSE ($\epsilon$). This measure takes into account the value of the prediction relative to the range of the variable variation. The quantity $\epsilon$ is a more representative value of discrepancy between predictions and test data than the RMSE itself because it is relative to the data. This means that the value is independent of each particular variable, and it can be used to compare the emulator performance in different variables easily. Equation~\ref{eq:epsilon} shows explicitly how $\epsilon$ is calculated. 
    
    \begin{equation} \label{eq:epsilon}
    \epsilon =\frac{1}{N} \sqrt{\sum_{i=1}^{N} \left( \frac{\hat{y}_i - y_i}{\max \mathbf{y} - \min \mathbf{y}} \right)^2 }
    \end{equation}
    where $N$ is the number of predictions considered, $y_i$ is the data at the $i$th point with $i=1,2,\dots,N$, $\hat{y}_i$ is the machine learning prediction of $y_i$, and $\mathbf{y}=\left\lbrace y_1, y_2, \dots, y_N\right\rbrace $.
    
\begin{figure}[!htbp] 
        \centering
        \subfigure[$\epsilon$ evaluated on the 30 test data points for the length of the longest crack.\label{fig:RMSE1}]
        {\includegraphics[width=0.44\columnwidth]{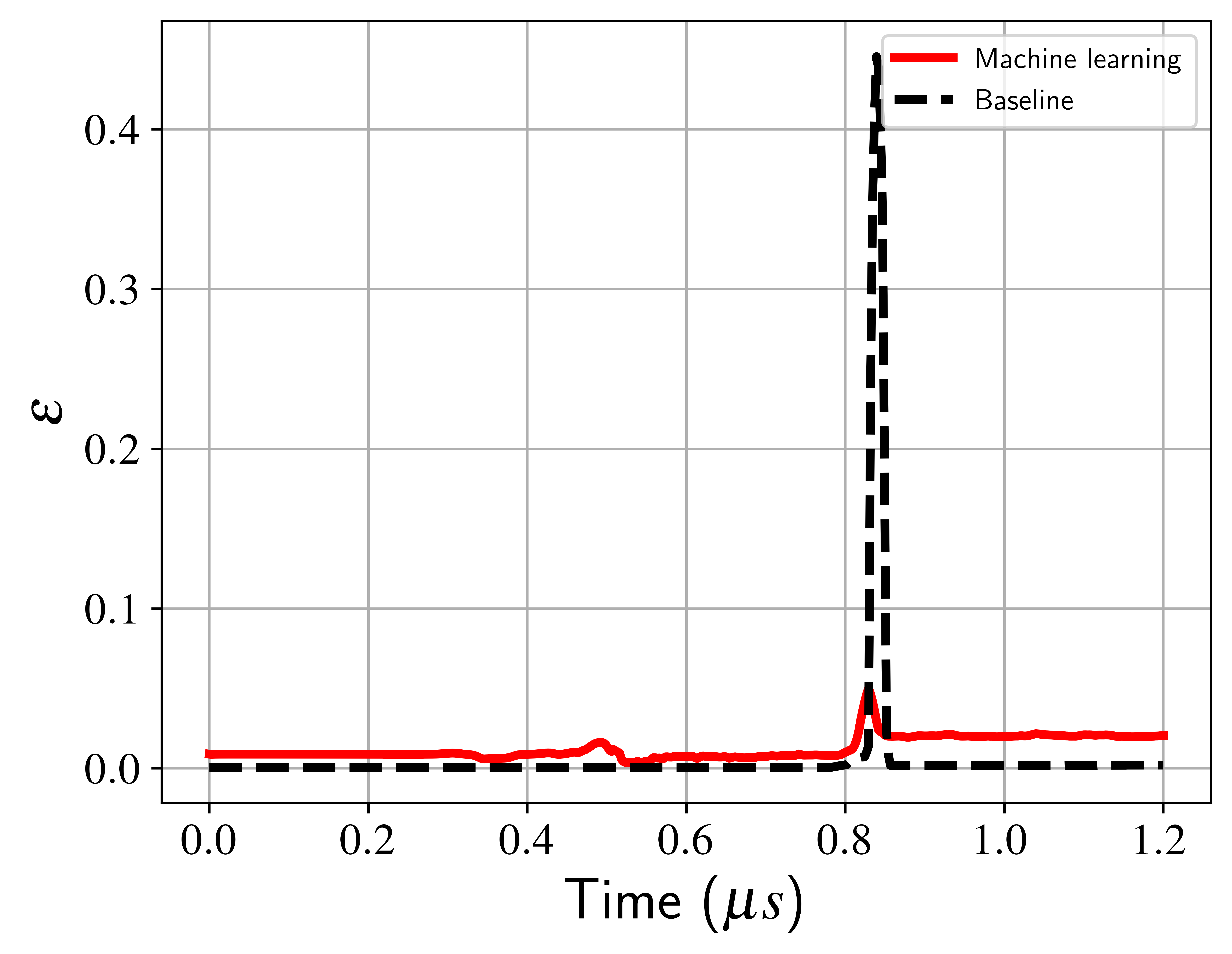} } 
        \hspace{2em}%
        \subfigure[$\epsilon$ evaluated on the 30 test data points for the maximum tensile stress.\label{fig:RMSE2}]
        {\includegraphics[width=0.46\columnwidth]{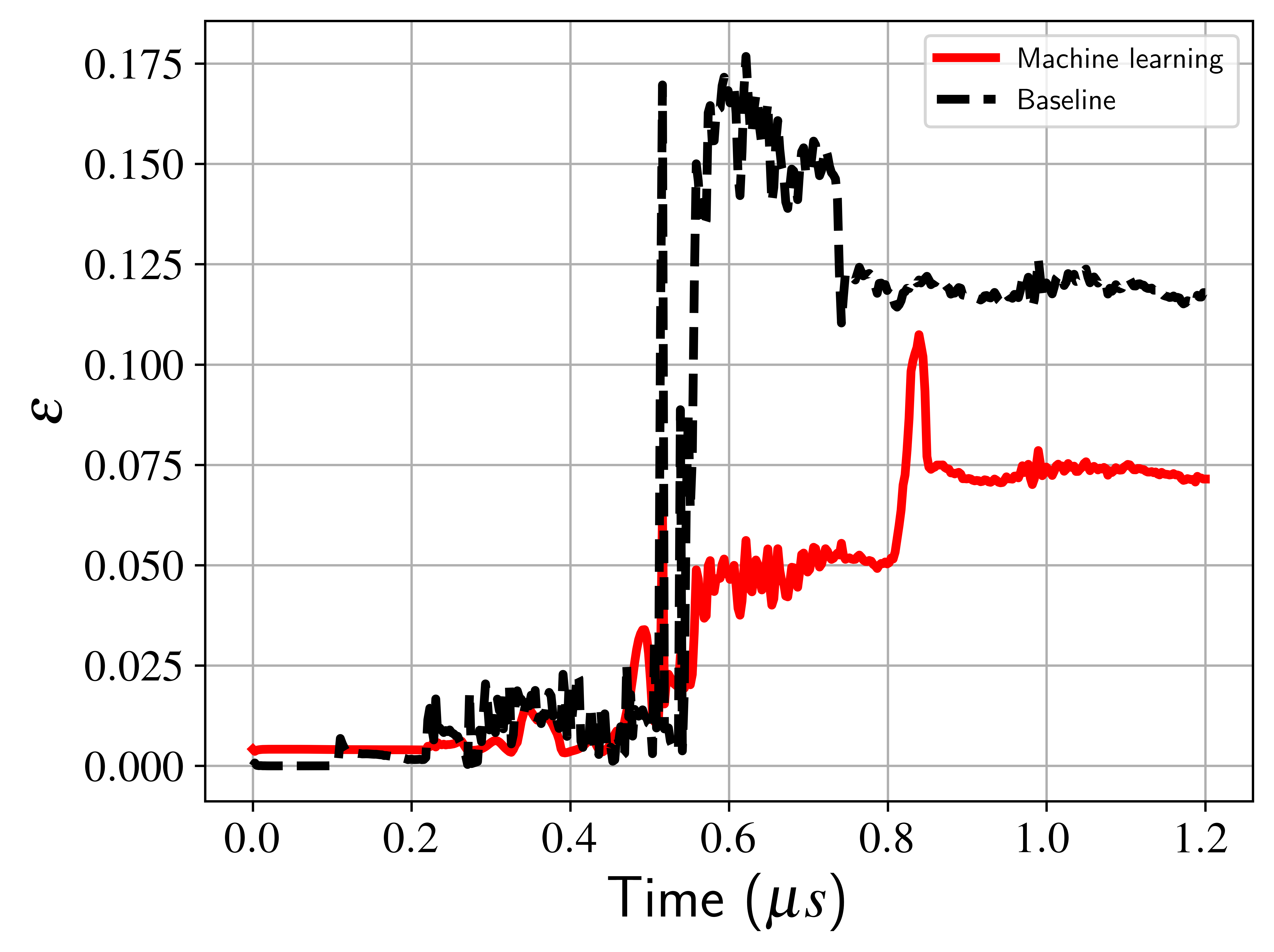} } 
        \caption{ $\epsilon$ as a function of time calculated using the 30 test data points. In black is shown $\epsilon$ of the baseline emulator and in red $\epsilon$ of the machine learning emulator. The machine learning emulator outperforms the baseline emulator where the data variation is higher. \label{fig:RMSE}}
    \end{figure}

    Figure~\ref{fig:RMSE} shows $\epsilon$ for the length of the longest crack and the maximum tensile stress comparing the error of the baseline emulator (Eq.~\ref{eq:mean}) against the error of the machine learning emulator. Figure~\ref{fig:RMSE} shows in black $\epsilon$ of the baseline emulator and in red $\epsilon$ associated with our machine learning emulator. Figure~\ref{fig:RMSE1} shows $\epsilon$ for the length of the longest crack as a function of time and Figure~\ref{fig:RMSE2} for the maximum tensile stress. The challenge in training the length of the longest crack is the steep jump in this variable at the time of failure (approximately 0.8 $\mu s$), which is where lies most of the variability (see Figure~\ref{fig:100Sim1}). Before and after this time, the length of the longest crack is almost constant, and it has almost no variability between simulations. As we see in Figure~\ref{fig:RMSE1}, the machine learning emulator significantly outperforms the baseline emulator around that critical time. Similarly, the challenge in training the maximum tensile stress is in the time range after the material failure (0.8 $\mu s$ to 1.2 $\mu s$), this is where most of the variability occurs between simulations (see Figure~\ref{fig:100Sim2}). As we see in Figure~\ref{fig:RMSE2}, the machine learning emulator also outperforms the baseline emulator in this range.
    
    \clearpage 
    
    \subsection{Machine Learning Validation} \label{sec:val}
    
\begin{figure}[ht!] 
        \centering
        \includegraphics[width=0.8\columnwidth]{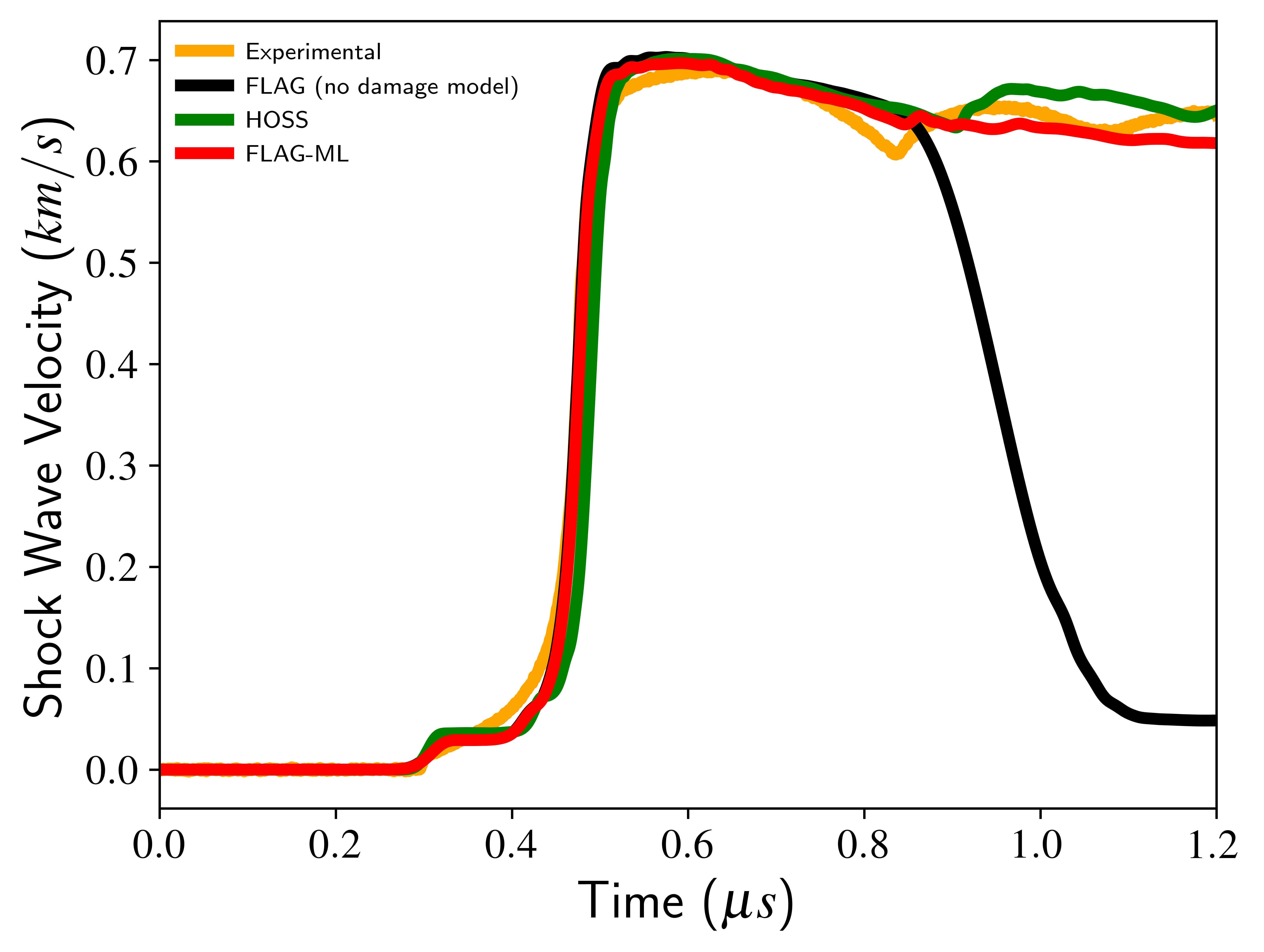} 
        \caption{Evolution of the shock wave velocity at the middle rear of the target plate for an experiment of impact velocity of 0.721$km/s$. FLAG-ML improves remarkably the predictions given by FLAG obtained without an imposed damage model.}
        \label{val2}
    \end{figure}

    We use the machine learning emulator trained with HOSS to inform the FLAG damage model. As Figure~\ref{val2} shows, using machine learning (FLAG-ML) captures the rebounding shock wave as HOSS does, improving remarkably the predictions given by FLAG without a damage model.
    
\begin{figure}[ht!] 
        \centering
        \includegraphics[width=0.8\columnwidth]{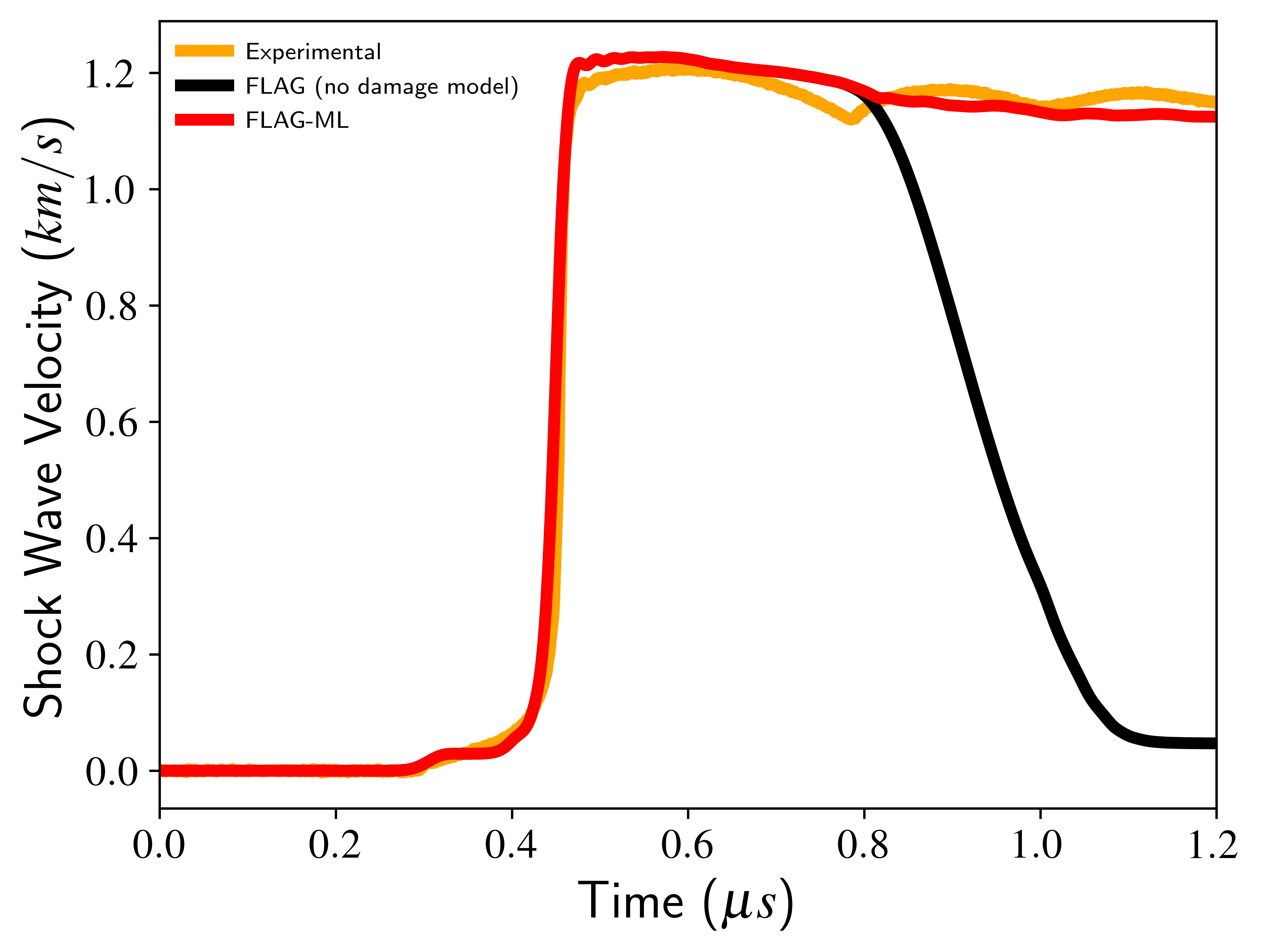} 
        \caption{Evolution of the shock wave velocity at the middle rear of the target plate for a second experiment with an impact velocity of 1.246 $km/s$. FLAG-ML improves the predictions given by FLAG obtained without an imposed damage model.}
        \label{val3}
    \end{figure}

    To test our FLAG-ML performance outside the training window, we have also compared it with experiments where the flyer plate impact velocity was increased from 0.721$km/s$ to 1.246$km/s$ (experiment 55-430-P30 in Cady el al.~\cite{cady2011alamos}). Besides the change in impact velocity, this second experiment was identical to the one described in Section~\ref{sec:HOSS}. The machine learning emulator used was the one described in Section~\ref{sec:ML}, and the training data used came from the HOSS simulations described in Section~\ref{sec:HOSS} (which all had an impact velocity of 0.721$km/s$, so this validation is outside the training window). Figure~\ref{val3} shows the experimental data, the FLAG (no damage model), and the FLAG-ML predictions for this new experiment. As we can see, FLAG-ML predictions are very close to the experimental data and significantly improve upon the FLAG (no damage model) predictions even outside the training window. Larkin et al., 2020~\cite{larkin2020scale} also show good agreements for this higher velocity experiment, however in this work, we use a machine learning emulator that allows reducing the computation time by four orders of magnitude.
    
    This proves that sufficient material-specific crack statistics were available to accurately approximate the results of simulations with similar loading conditions. This may reduce the need for further high-fidelity simulations.
    
    \section{Conclusion}\label{sec:conclusion}
    
    The main contribution of this work is to develop a machine learning workflow for developing more accurate yet efficient materials models by bridging scales. This workflow enables us to model materials undergoing brittle failure more from first principles physics by abstracting explicit phenomena like crack evolution and coalescence. The resulting workflow allows for accuracy comparable to a high fidelity model (such as HOSS) but in a fraction of the time, achieving orders of magnitude speed-up and allowing for the real-time embedding of crack statistics propagation in a continuum model (such as FLAG). The workflow developed here is agnostic to the actual codes used, because the underlying machine learning model only requires crack statistics generated from any mesoscale model and translates those to effective elastic moduli in a continuum scale model. 
    
    Machine learning is a powerful tool for dimension reduction. Our algorithm was able to identify the key physics features that were responsible for driving the damage degradation at the continuum scale. Our approach relied on data from HOSS simulations to train an appropriate machine learning emulator. The machine learning emulator was then be used to inform FLAG simulations in a computationally efficient manner. This efficiency is possible because the machine learning emulator is $\approx 10^4$ times faster than HOSS (while retaining comparable accuracy). This speed-up is especially critical when the initial crack distribution is inherently stochastic, and many simulations are often required to estimate predictive uncertainty or perform calibration.
    
    We validated the combination of machine learning and FLAG on two flyer plate experiments with velocities of 0.7 km/s and 1.2km/s. While the machine learning algorithm was trained at the lower velocity, it performed with high accuracy while validating against the second experiment, which had a 40\% higher flyer plate impact velocity.
    
    The ability to transfer learn crack statistics between the two impact velocities gives us reasonably high confidence in using this workflow for predicting any scenarios with impact velocities between the two experimental values. This is the first time such a workflow has been shown to work successfully in comparison to experiments. In future work, we aim to explore the performance of this approach under more diverse loading conditions.
    
    \section*{Acknowledgments}
    
    This work was performed under U.S. Government contract 89233218CNA000001 for Los Alamos National Laboratory (LANL), which is operated by Triad National Security, LLC for the U.S. Department of Energy/National Nuclear Security Administration. Approved for public release LA-UR-20-22890.
    
    The authors gratefully acknowledge support from the Advanced Simulation and Computing (ASC) Program.
    
    \section*{Data Availability}
    
    Available from the authors upon request.
    
    \bibliographystyle{spmpsci}
    \bibliography {bibliography}
    
\end{document}